\documentclass[prb,twocolumn]{revtex4-1}
\usepackage[table, dvipsnames]{xcolor}
\usepackage{graphicx}
\usepackage{amssymb}
\usepackage{dcolumn}
\usepackage{float}
\usepackage{bm}
\usepackage{multirow}
\usepackage{booktabs}
\usepackage[T1]{fontenc}
\usepackage[utf8]{inputenc}
\usepackage{lmodern}
\usepackage{color}
\usepackage{makecell}
\usepackage{tikz}
\usepackage{pgfplots}
\usepackage{mathtools}
\usepackage{physics}
\usepackage{amsmath}
\usepackage{setspace}
\usepackage{ragged2e}

\usepackage[colorlinks=true,linkcolor=red]{hyperref}
\newcommand\scalemath[2]{\scalebox{#1}{\mbox{\ensuremath{\displaystyle #2}}}}

\newcommand{\mathleft}{\@fleqntrue\@mathmargin0pt}

\newcolumntype{M}[1]{>{\centering\arraybackslash}m{#1}}

\DeclareMathAlphabet{\bi}{OML}{cmm}{b}{it}

\def\be{\begin{equation}}
	\def\ee{\end{equation}}
\def\bearr{\begin{eqnarray}}
	\def\eearr{\end{eqnarray}}

\begin{document}
	
	\title{Floquet-Engineered Valley-Topotronics in Kekul$\'e\ $- Y Bond Textured Graphene Superlattice}
	\bigskip
	\author{Sushmita Saha and Alestin Mawrie}
	\normalsize
	\affiliation{Department of Physics, Indian Institute of Technology Indore, Simrol, Indore-453552, India}
	\date{\today}
	\begin{abstract}
	The exquisite distortion in a Kekul$\'e\  $-Y (Kek-Y) superlattice merges the two inequivalent Dirac cones (from the $K$- and the $K^\prime$- points) into the highest symmetric $\Gamma$-point in the hexagonal Brillouin zone. Here we report that a circularly polarised light not only opens up a topological gap at the $\Gamma$-point but also lifts the valley degeneracy at that point. Endowed with Floquet dynamics and by devising a scheme of high-frequency approximation, we have proposed that the handedness (left/right) in polarised light offers the possibility to realize the valley-selective circular dichroism in Kek-Y shaped graphene superlattice. Also, the non-vanishing Berry curvature and enumeration of valley resolved Chern number $\mathcal{C}_{K}/\mathcal{C}_{K^\prime}=+1/-1$ enable us to assign two pseudospin flavors (up/down) with the two valleys. Thereby, the above observations confirm the topological transition suggesting the ease of realising the valley quantum anomalous Hall (VQAH) state within the photon-dressed Kek-Y. 
	%Furthermore, we have established that the optical valley selection rule is obeyed. 
These findings further manifest a non-zero optical valley polarisation which is maximum at the $\Gamma$-point. %with respect to the left and right circularly polarized light, respectively. %, which exponentially reduces at points away from the $\Gamma$-point. %As the k-value moves away from the high-symmetric $\Gamma$-point, the degree of polarization gradually diminishes towards zero.
	Our paper thus proposes an optically switchable topological valley filter which is desirous in the evolving landscape of valleytronics. %Concisely, our work promises an approach to engineer and utilize the electronic valley degrees of freedom.
	\end{abstract}
	
	\email{amawrie@iiti.ac.in}
	
	\maketitle
	
	\section{Introduction}
\begin{figure}[b]
	\includegraphics[width=87.5mm,height=45.5mm]{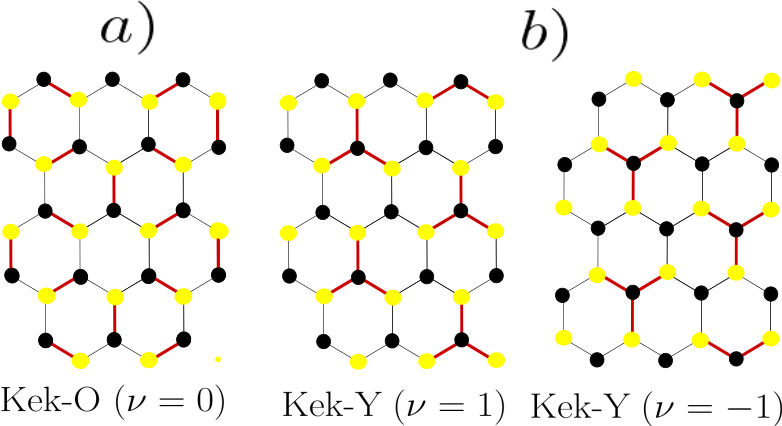}
	\includegraphics[width=42.5mm,height=22.5mm]{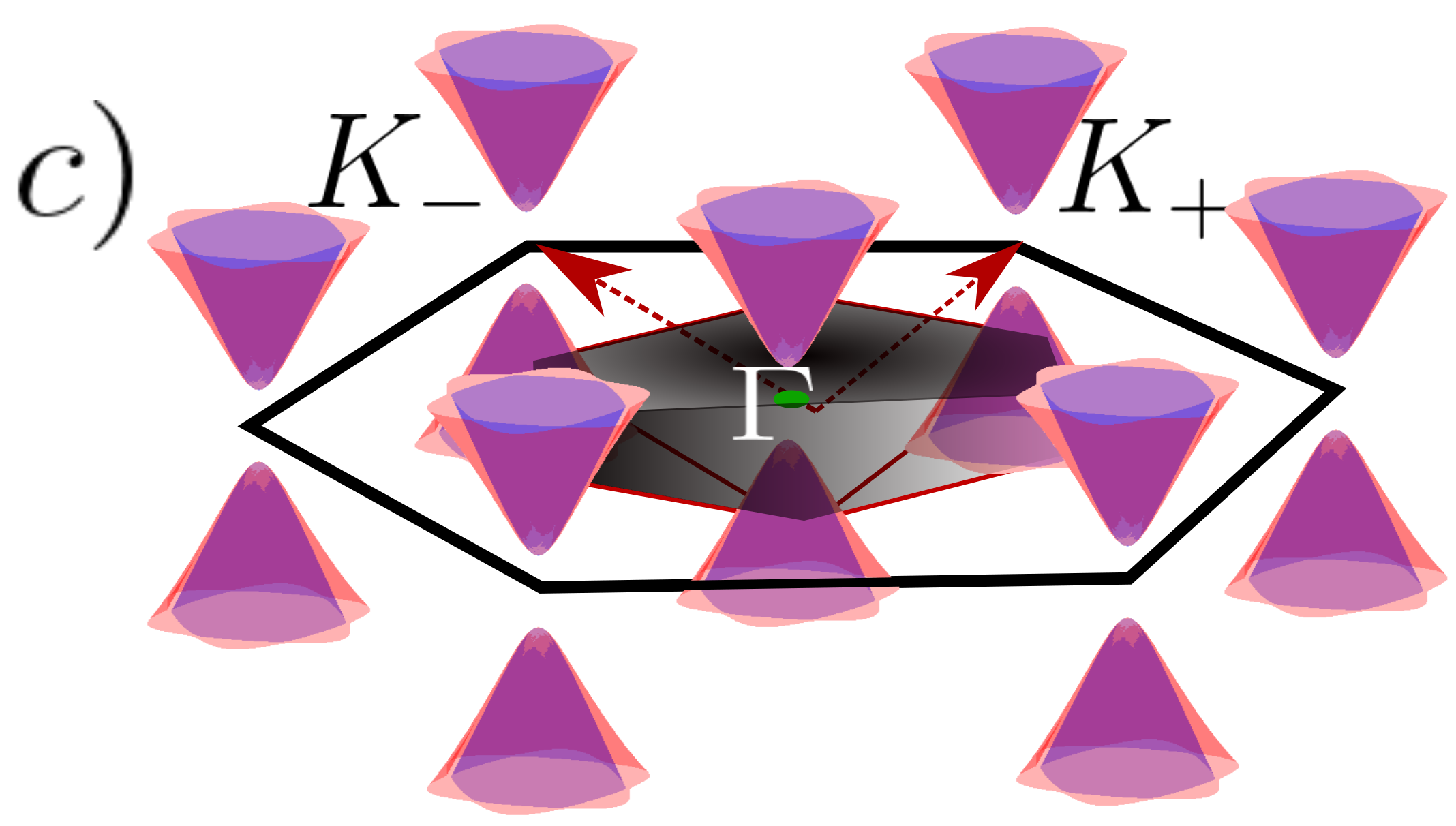}
	\includegraphics[width=42.5mm,height=22.5mm]{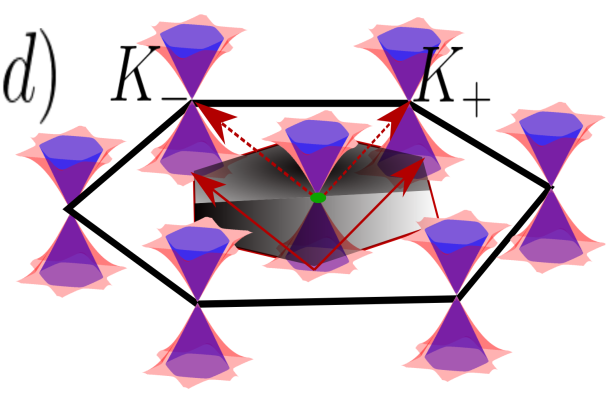}
	\caption{Schematic of honeycomb superlattices with (a) Kek-O and (b) Kek-Y shaped bond distortion. Reciprocal lattice of (c) a Kek-O type and (d) a Kek-Y bond textured graphene superlattice, modified by two inequivalent Dirac cones folded onto the high symmetric $\Gamma$-point in the Kek-Y distorted lattice. The red arrows represent the reciprocal lattice vectors ($\textbf{K}_\pm=\frac{2\pi}{9}\sqrt{3}(\pm1,\sqrt{3})$) in the reduced Brillouin zone. Whereas the Kek-O lattice shows a gap dispersion at all symmetric $\Gamma$-points, the dispersion of Kek-Y lattice is gapless.}
	\label{Fig1}
\end{figure}

%In the emerging sphere of technology, an intense focus has been drawn towards the need of achieving faster, highly efficient and low energy consumed electronic devices. 
The increasing need of data storage or computer logic systems endeavours the condensed matter research to exploit all possible degree of freedom (DOF) in a quantum material. One such DOF's are the ``valley-pseudospins'' giving rise to the valleytronics field
\cite{valley1,valley2,valley3,valley4}. In a valleytronic material, we look for a scope to exploit the carrier's valley DOF for the purpose of encoding quantum information. %, analogous to charge in electronics or spin in spintronics.
The conduction band in a 2D hexagonal structure such as graphene disperses local minima termed as Dirac ``valley's'' (the $K$ and $K^\prime$ points). These are two energy degenerate inequivalent points in the Brillouin zone. 
The exquisite distortion in a Kekul$\'e\  $-(Y \& O)-shaped (which we will refer to as Kek-Y and Kek-O, respectively) hexagonal lattice compels the two inequivalent Dirac valleys to merge\cite{merge1,merge2,merge3,kekbasic,Gamayun} onto the highly symmetric $\Gamma$-point [\ref{Fig1} c) \& d)]. %{(unlike the regular hexagonal lattice, where the valley degenracy is occuring at the $K_+$ and $K_-$ points in the Brillouin zone}). 
The first Brillouin zone of the distorted lattice would then looks like in Fig. [\ref{Fig1} c) \& d)], whose area is shrunk to 1/3$^{\rm rd}$ of the Brillouin zone of undistorted lattice. The experimental realization of a distorted Kek-Y (shown in Fig. [\ref{Fig1} b)]) lattice structure in 2D graphene superlattice has been confirmed by Gutiérrez et al\cite{exptkekY, exptkekY1}. The other possible distortion (shown in Fig. [\ref{Fig1} a)]) identified as a distorted Kek-O lattice is however not experimentally feasible{\cite{KekO1,KekO2}}.

It can be shown that the distortion that brings about the Kek-Y lattice is not accompanied by any breaking of the time reversal symmetry(TRS), implying that the $\Gamma$-point in the Brillouin zone of Fig.[\ref{Fig1} d)] is doubly degenerate with the two valleys. Since the valley degeneracy occurs at the same $\Gamma$-point in the Brillouin zone, the Kek-Y graphene superlattice offers a better scope in manipulating the two valleys for the valleytronics application. This paper thus explores the possibility to manipulate the two valleys in a Kek-Y lattice at the $\Gamma$-point by creating non-equilibrium fermionic imbalance between them. {To execute the manipulation, we have used the circularly polarized light (CPL) as an external stimulus effectively creating a topological gap (at $\Gamma$-point in the Brillouin zone) by breaking the time reversal symmetry\cite{kane1,kane2}. Interestingly, we have also observed that this is accompanied by the lifting of the valley degeneracy at the same point. In the light of the Floquet dynamics\cite{kek,kekrad,floquetgr}, we have investigated the valley responses to the left/right circularly polarized light (CPL) in the THz frequency domain, and observe the valley-selective Chern number is either $\mathcal{C}_{K}$=+1 or $\mathcal{C}_{K^\prime}$=-1, for the $K$ and $K^\prime$-valley, respectively. This observation indicates progress towards achieving valley quantum anomalous Hall (VQAH) states. 
Subsequently, we show in this paper that the circular dichroism (CD)\cite{optichelicityMoS,opticpumpMoS} can also be achieved in a photon dressed Kek-Y superlattice. These investigations lead us to the immensely interesting valley-selective CD which is exact opportune for valleytronics applications as an optically switchable topological valley filter.
 	
This paper is further organised as follows. In Sec-\ref{sec2}, using the Floquet dynamics\cite{magnus,superfluid,prr1,prr2}, we provide the detailed formulation of the dynamic quasi-energy gaps in a Kek-Y distorted lattice. %the dispersion creating dynamical periodic gaps. 
Invoking high frequency approximation, in Sec-\ref{sec3}, we have detailed the existense of valley-contrasted CD. Sec-\ref{sec4} unfolds the degree of valley resolved optical polarisation. In Sec-\ref{sec5}, we consider a nanoribbon geometry to formulate the real space Floquet Hamiltonian of the Kek-Y lattice 
which guides to reveal the topological edge states\cite{book}.
We summarize the paper in Sec-\ref{sec6}. 

\section{Ground state energy of a Kek-Y lattice}\label{sec2}
To start with, our analysis of the Kek-(Y \& O)  zigzag-graphene monolayer in Fig. [\ref{Fig1} a) \& b)] employs the tight binding Hamiltonian given as\cite{Gamayun}
	\begin{eqnarray}
	\label{Eq:1}
		\hat{H}_\mathcal{TB}=-\sum_{\textbf{r}}\sum_{\xi=1}^{3} t_{{\bf r}, \xi} \;a^\dagger_\textbf{r}\,b_{\textbf{r}+\textbf{s}_\xi}+h.c.
	\end{eqnarray} 
Here $a_{\bf r}$ and $b_{{\bf r}+{\bf s}_\xi}$ are the creation and anihilation operators at the A (${\bf r}=p\boldsymbol{\delta}_1+q\boldsymbol{\delta}_2$, where $p,q\in \mathbb{Z}$) and B (${\bf r}+{\bf s}_\xi$) sublattice sites, respectively. The vector $\textbf{s}_1=\frac{1}{2}(\sqrt{3},-1)$, $ \textbf{s}_2=-\frac{1}{2}(\sqrt{3},1)$ and $\textbf{s}_3=(0,1)$, taking a unit lattice parameter. A usual hexagonal lattice such as graphene is illustrated by the hopping parameter $t_{{\bf r}, \xi}=t_0\approx2.7$ eV that gives rise to a dispersion $E({\bf k})=\pm \vert \epsilon(k)\vert$, defining $\epsilon({\bf k})=t_0\sum_{\xi=1}^3e^{-i{\bf k}\cdot {\bf s}_\xi}$. In a Kekul$\'e\  $ distorted lattice, the bond strength between the A and B-sublattices is modified as\cite{Gamayun}$t_{{\bf r},\xi}=t_0[1+2\,\text{Re}(\Delta e^{i(u\textbf{K}_\textbf{+}+v\textbf{K}_\textbf{-}).\textbf{s}_{\xi}+i\textbf{G}.\textbf{r}})]
$, where  $\textbf{K}_\pm=\frac{2\pi}{9}\sqrt{3}(\pm1,\sqrt{3})$ are the reciprocal lattice vectors of the reduced Brillouin zone (shown as red arrows in Fig. [\ref{Fig1} c) \& d)]). The Kekul$\'e\  $ wave vector ${\bf G} = {\bf K}_+ - {\bf K}_-$ couple the two Dirac valleys (at the $K,\; \& \;K^\prime$ points) with a complex coupling strength $\Delta=\Delta_0 e^{i 2\pi (u+v)/3}$ known as Kekul$\'e\  $parameter with amplitude $\Delta_0\leq0.1$. Here integers $(u,v)$ are the indices which generates the different Kekul$\'e\  $ textures accordingly to the relation ($\nu=1+v-u$ mod 3). Specifically, $\nu=0$ defines the Kek-O distorted phase and $\nu=\pm 1$ describes Kek-Y graphene distorted phases (Fig. [\ref{Fig1} a) \& b)]). To understand the low energy bands of this Kekul$\'e\ $graphene superlattice, we transform Eq. (\ref{Eq:1}) into the momentum space as

%{\color{red}The $K$ and $K^\prime$ valleys (at the green Dirac points) are coupled by the wave vector ${\bf G} = {\bf K}_+ - {\bf K}_-$ of the Kekule bond texture with a complex coupling strength $\Delta=\Delta_0 e^{i 2\pi (u+v)/3}$ known as Kekul$\'e\  $parameter  with amplitude $\Delta_0\leq0.1$. The integers $(u,v)$ are the indices that generates a Kekule texture accordingly to the relation given below 
%\begin{equation}
%\nu=1+v-u \text{ mod }3,
%\end{equation}	
%with $\nu=0$ describing the Kek-O distorted phase and $\nu=\pm 1$ the kek-Y graphene distorted phases (Fig. [\ref{Fig1} a) \& b)]).} %and their  Resultant is considered as Kekul$\'e\ $ wave vector i.e. $\textbf{G}=\frac{4\pi}{9a}\sqrt{3}(1,0)$. The coupling strength between the $K$ and $K^\prime$ points is captured by the parameter   .
 %perform merging of the Dirac points into the highly symmetric $\Gamma$ point as shown in Fig[\ref{Fig1}(b)]. 

	\begin{eqnarray*}
		\hat{H}_{0}\textbf{(k)}&&=-\epsilon\textbf{(k)}a_\textbf{k}^\dagger\,b_\textbf{k}-\Delta\,\epsilon\textbf{(k}+u\textbf{K}_++v\textbf{K}_-\textbf{)}a_\textbf{k+G}^\dagger b_\textbf{k}\nonumber\\&&-\Delta^\ast\epsilon \textbf{(k}-u\textbf{K}_+-v\textbf{K}_-\textbf{)}a_{{\bf k}-{\bf G}}^\dagger b_\textbf{k}+h.c.
	\end{eqnarray*}
Using the relations $\epsilon({\bf k})=\epsilon({\bf k}+3{\bf K}_\pm)=e^{2\pi i/3}\epsilon({\bf k}+{\bf K}_++{\bf K}_-)$, the above Hamiltonian can be written in a $6\times 6$ matrix form, with respect to the basis vector
$c_\textbf{k}=\begin{pmatrix}
a_\textbf{k} & a_{\textbf{k}-\textbf{G}} &a_{\textbf{k}+\textbf{G}} &b_\textbf{k} & b_{\textbf{k}-\textbf{G}} & b_{\textbf{k}+\textbf{G}}
\end{pmatrix}^T$, such that
\begin{eqnarray}\label{6by6}
\begin{aligned}
\hat{H}_0(\textbf{k})=-c_\textbf{k}^\dagger
\begin{pmatrix}
	0_{3\times 3} & \hat{\Sigma}_\nu(\textbf{k})\\
\hat{\Sigma}_\nu^{\dagger}(\textbf{k}) & 0_{3\times 3} 
\end{pmatrix} c_\textbf{k}.
\end{aligned}\end{eqnarray}
Here $0_{3\times 3}$ is a null matrix and $\hat{\Sigma}_\nu({\bf k})$ is a $3\times 3$ matrix given as following
		\begin{align*}
		\hat{\Sigma}_\nu({\bf k})=\begin{pmatrix}
			\epsilon_0({\bf k}) & \Delta \epsilon_{\nu+1}({\bf k}) & \Delta^\ast \epsilon_{-\nu-1}({\bf k}) \\\Delta^\ast \epsilon_{1-\nu}({\bf k}) & \epsilon_{-1}({\bf k}) & \Delta \epsilon_{\nu}({\bf k})\\\Delta \epsilon_{\nu-1}({\bf k}) & \Delta^\ast \epsilon_{-\nu}({\bf k}) & \epsilon_1({\bf k})
		\end{pmatrix},
	\end{align*} 
defining $\epsilon_\nu=\epsilon(\textbf{k}+\nu\textbf{G})$. 

We now check the nature of the dispersion with respect to the Hamiltonian in Eqn [\ref{6by6}] for $\nu=\pm 1$ (the Kek-Y distorted phase). Instead of the Dirac cones at the usual valley points of the reciprocal hexagonal lattice, the exquisite Kek-Y distortion causes the Dirac conical singularity to appear at the $\Gamma$-point as well bringing out a reduced first Brillouin zone (shown in the shaded area of Fig.[\ref{Fig1} c) \& d)] shrunk by 1/3$^{\rm rd}$ of the hexagonal Brillouin zone in graphene.
There are four low energy bands with respect to the above Hamiltonian in Eq. [\ref{6by6}] as shown in Fig. [\ref{Fig1} d)]).  %Symmetric points including the $K$, $K^\prime$ and the $\Gamma$-points now feature a double Dirac cones and they act as equivalent points in the Brillouin zone. Several literature descibes this as the folding of the Dirac cones from the $K$ and  $K^\prime$ points into the $\Gamma$-point. 
 %The figure of the dispersions arising from the $6\times 6$ Hamiltonian about a direction $\Gamma-K^\prime-K-\Gamma-K^\prime$ is shown in Fig. [\ref{label}].
%To emphasise the dispersion near the $\Gamma$-point, we have managed to eliminate two higher energy bands. 
In the vicinity of the $\Gamma$-point, the energy spectrum is governed by $\mathcal{\psi}_{\bf k} = \begin{pmatrix}
a_{{\bf k} - {\bf G}} & a_{{\bf k} + {\bf G}} & b_{{\bf k} - {\bf G}} & b_{{\bf k} + {\bf G}}
\end{pmatrix}^T$. When projected onto this subspace, Eq. [\ref{6by6}] reduces to $\hat{H}_0({\bf k})=\mathcal{\psi}_{\bf k}^\dagger \hat{\mathcal{H}}_0({\bf k})\mathcal{\psi}_{\bf k}$, such that \begin{align}\label{4by4}
\scalemath{0.98}
{\hat{\mathcal{H}}_0(\textbf{k})=
	\begin{pmatrix}
	0 & 0 & \epsilon_{-1}(\textbf{k}) & \tilde{\Delta}\epsilon_{\nu}(\textbf{k})\\
	0 & 0 &\tilde{\Delta}^\ast\epsilon_{-\nu}(\textbf{k}) & \epsilon_{1}(\textbf{k})\\
	\epsilon_{-1}^\ast(\textbf{k}) & \tilde{\Delta}\epsilon_{-\nu}^\ast(\textbf{k}) & 0 & 0\\
	\tilde{\Delta}^\ast\epsilon_{\nu}^\ast(\textbf{k}) & \epsilon_{1}^\ast (\textbf{k}) & 0 & 0
	\end{pmatrix}}.
\end{align}

%{\color{red}The low-energy spectrum is governed by the four modes $\mathcal{K}_{\bf k} = \begin{pmatrix}
%a_{{\bf k} - {\bf G}} & a_{{\bf k} + {\bf G}} & b_{{\bf k} - {\bf G}} & b_{{\bf k} + {\bf G}}
%\end{pmatrix}^T$, which for small ${\bf k}$ lie near the Dirac points at $\pm {\bf G}$ . (We identify the $K$ valley with $+ {\bf G}$ and the $K^\prime$ valley with $- {\bf G}$.) %When projected onto this subspace, Eq. [\ref{6by6}] reduces to $H_0({\bf k})=\mathcal{K}_{\bf k}^\dagger \mathcal{H}_0({\bf k})\mathcal{K}_{\bf k}$, such that
%\begin{align}\label{4by4}
%\scalemath{0.97}
%{\mathcal{H}_0\textbf{(k)}=
%\begin{pmatrix}
%0 & 0 & \epsilon_{-1}(\textbf{k}) & \tilde{\Delta}\epsilon_{\nu}(\textbf{k})\\
%0 & 0 &\tilde{\Delta}^\ast\epsilon_{-\nu}(\textbf{k}) & \epsilon_{1}(\textbf{k})\\
%\epsilon_{-1}^\ast(\textbf{k}) & \tilde{\Delta}\epsilon_{-\nu}^\ast(\textbf{k}) & 0 & 0\\
%\tilde{\Delta}^\ast\epsilon_{\nu}^\ast(\textbf{k}) & \epsilon_{1}^\ast (\textbf{k}) & 0 & 0
%\end{pmatrix}}.
%\end{align}	
%}

To understand the Dirac nature of the low-energy dispersion, we expand the above Hamiltonian about ${\bf k}=0$. %which in this case is about  any of the high symmetric points in the Brillouin zone (say the $\Gamma$-point). %As the $\Gamma$-point asserts a high symmetric point due to the folding of the $K$ and $K^\prime$ valleys, the low energy approximation about the $\Gamma$-point is significant. 
The above Hamiltonian when Taylor expanded about such points yields a 4-component Dirac equation of the form
	\begin{align}\label{Eq4}
\hat{\mathcal{H}}_{0}\textbf{(k)}={\hbar} [v_0 ({\bf k}\cdot \boldsymbol{\sigma})\otimes \tau_0+\Delta v_0 \sigma_{0}\otimes({\bf k}\cdot \boldsymbol{\tau})].
	\end{align}
%	$k_\pm=\textbf{k}_x\pm i \textbf{k}_y$ is written interms of momentum $\textbf{k}=(k_x,k_y)$, 
Here, $\boldsymbol{\sigma}=(\sigma_x,\sigma_y)$ stands for Pauli matrices and ${\sigma}_0$ is the identity matrix. The second set of Pauli's matrices $\boldsymbol{\tau}=(\tau_x,\tau_y)$ and $\tau_0$ act on the valley degree of freedom. We have also defined the Fermi velocity $v_0=\frac{3}{2}t_0 a_0/\hbar$. As a result of the Kek-Y distortion in the lattice, the low $k$- approximation leads to two concentric Dirac cones with eigenenergies \begin{eqnarray*}
	\label{Eq5}
		E_{\lambda\gamma}(k)=\gamma\hbar v_0 k(1+\lambda \Delta_0)
	\end{eqnarray*} 
	The folding of $K,K^\prime$ onto the $\Gamma$-point insists to assign the cone index($\lambda$) and the band index($\gamma$) with the energy levels as 
 $\lambda=+/-$ denotes the external/internal cones and band index $\gamma=+/-$ the conduction/valance band. Clearly, the two valley bands are degenerate at the $\Gamma$-point, indicating that there are no topological phase transition due to the Kek-Y distortion. Here the motivation comes to investigate how the band diagram evolves if the TRS is broken by the application of CPL.  
\begin{figure}[b]
		\includegraphics[width=65.5mm,height=40.5mm]{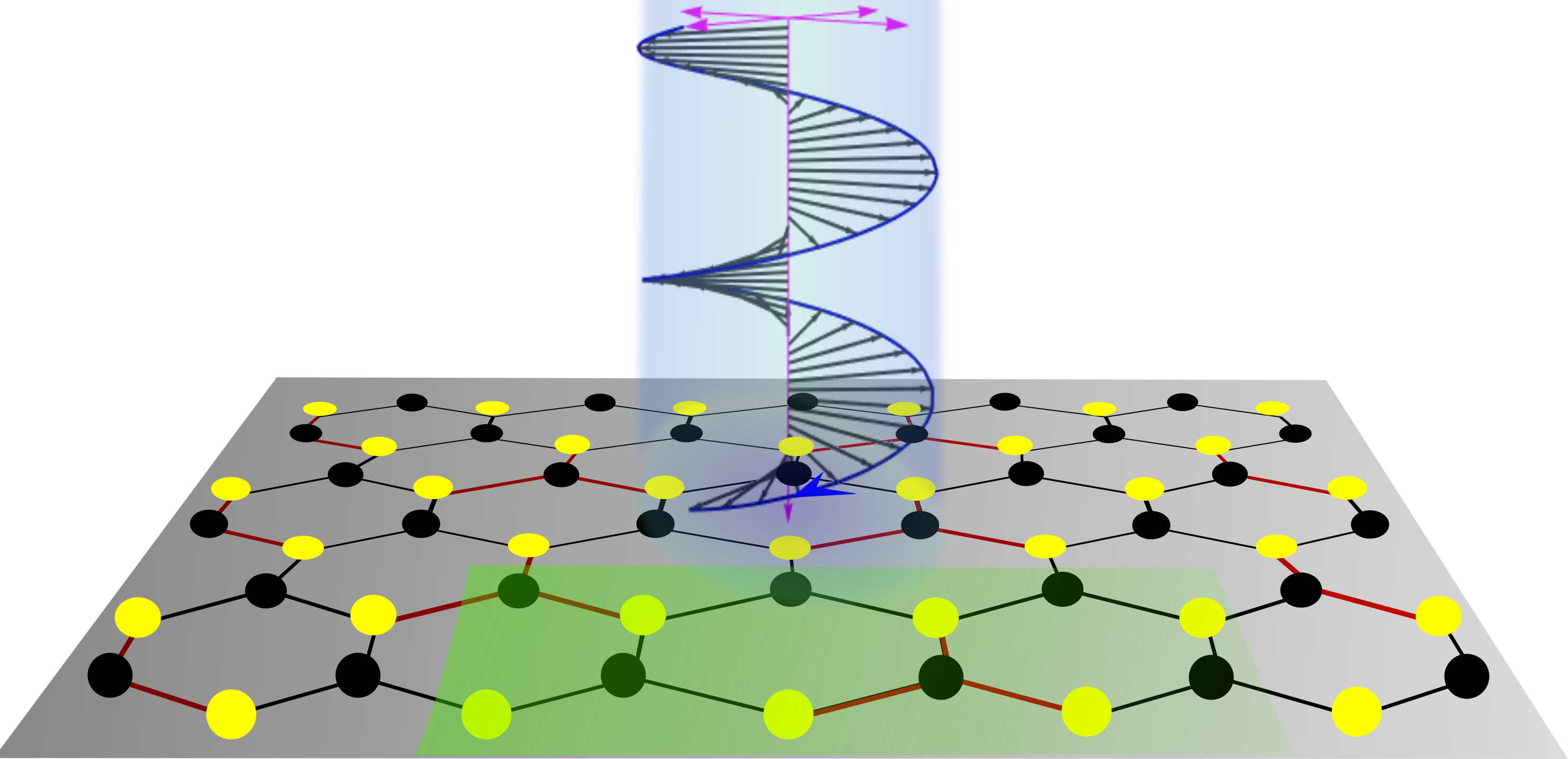}
		\caption{The schematic represents the interaction between Kek-Y bond ordered Graphene with the perpendicularly incident Circularly Polarized Light.\, Here Green shaded area indicates the unit cell of Kek-Y superlattice which happens to be three times larger than the Pristine Graphene one's.}
		\label{Fig2}
	\end{figure}
\section{Optically driven floquet formalism for Kek-Y distorted graphene }\label{sec3}
%\textit{\textbf{Construction of Floquet Hamiltonian in Momentum Space }}:
In this section, we will dig into the Floquet dynamics of Kekul$\'e\ $ distorted graphene. To drive the system into a topological phase, a CPL is shone perpendicularly to the plane of a 2D sheet as shown in Fig. [\ref{Fig2}]. %{\color{red}We assume that the dimension of the sample is smaller compared to the wavelength of light. }
Under the influence of this time-periodic and spatially homogeneous CPL, Kekul$\'e\ $ lattice is shaken by the vector potential $\textbf{A}(t)=A_0[\cos(\omega t),\zeta \sin(\omega t)]$ with amplitude $A_0={\varepsilon_0}/{\omega}$ ($\varepsilon_0$ is the magnitude of the effective electric field and $\omega$ is the frequency of the incident light) where  $\zeta$ denotes the handedness of the polarization such that $\zeta=+/-$ belongs to left/right-handed polarised light. The corresponding electric field experienced by the fermionic charge carriers is $\textbf{E}=-{\partial \textbf{A}}/{\partial t}=\varepsilon_0[\sin(\omega t) \;\;-\zeta \cos(\omega t)]$. 
	
To show that this time-periodic perturbation modifies the dispersion of a quantum particle moving in a spatially periodic lattice potential, we used Pierl’s substitution $\textbf{k}\rightarrow \textbf{Q}= \textbf{k}+e\textbf{A}(t)/\hbar$  and rewrite Eq. (\ref{Eq4}) as\begin{align}\label{Ra_H}
\hat{\mathcal{H}}_{0}(\textbf{k},t)=\hbar v_0
		\begin{pmatrix}\textbf{Q}\cdot \boldsymbol{\sigma}  & \Delta\,Q_{+} \boldsymbol{\sigma}_{0}  \\ \Delta^\ast\,Q_{-}\boldsymbol{\sigma}_{0} &\textbf{Q}\cdot \boldsymbol{\sigma} \end{pmatrix},
	\end{align}
where we have defined $Q_{\pm}=Q_{x}\pm i Q_{y}$. The solution  to the Schr$\ddot{o}$dinger equation with respect to the above Hamiltonian is 
\begin{eqnarray}
\vert\Psi_\alpha(t)\rangle=e^{-i E_\alpha t/\hbar}\vert \psi_\alpha(t)\rangle
\end{eqnarray} where the Floquet states $\psi_\alpha(t)$ are the solutions of the following differential equation
\begin{eqnarray}\label{key1}
\hat{\mathcal{H}}_F(t)\vert\psi_\alpha(t)\rangle=E_\alpha \psi_\alpha(t)\rangle,
\end{eqnarray} with $E_\alpha$ being the Floquet quasi-eigen energy states. We have defined the time-dependent Floquet Hamiltonian as $\hat{\mathcal{H}}_F(t)=\hat{\mathcal{H}}_0(\textbf{k},t)-i\hbar \partial_t$. % which acts on $\psi_\alpha(t)$. 
 Also,  %$\epsilon_n$ being the quasi-energy of the radiation driven system and 
$\vert\psi_\alpha(t)\rangle=\vert\psi_\alpha(t+T)\rangle$ 
has the same periodicity as the Hamiltonian in Eq. (\ref{Ra_H}) which allows us to write  $\vert\psi_\alpha(t)\rangle=\sum_{n=-\infty}^{^\infty}e^{i n \omega t}\vert\phi_\alpha^n\rangle$ as a solution of Eq. (\ref{key1}). The total wavefunction now live in the extended Floquet-Hilbert space ($\mathcal{F}=\mathcal{H}\otimes\mathcal{T}$), with $\mathcal{H}$ being the usual Hilbert space for a system under no radiation and $\mathcal{T}$ is the space formed by a complete set of periodic function $e^{in\omega t}$.  %The advantage of the Floquet theorem is that it provides the set of solutions of the time-depenndent Schrodinger equation, taking the solution as , with
% The quasi-energy eigenvalue problem is now shaped as \begin{eqnarray}
%		\hat{\mathcal{H}}_F(t)\psi_\alpha(t)=E_\alpha\psi_\alpha(t).
%	\end{eqnarray}
%In the given exted Hilbert space, one can define
%\begin{eqnarray}
%\frac{1}{T}\int_0^T\langle\psi_{n^\prime}(t)\vert\psi_n(t)\rangle dt=\delta_{nn^\prime }
%\end{eqnarray}
%It provides the Floquet modes not only at the initial state but also includes their full periodicity. 
Writing $\vert\phi_\alpha\rangle=\begin{pmatrix}
.&.&\vert\phi_\alpha^{-1}\rangle&\vert\phi_\alpha^{0}\rangle&\vert\phi_\alpha^{1}\rangle&.&.
\end{pmatrix}^T$ in the said Floquet-Hilbert space, one can arrive at the following equation  %the orthoonormality condition  from the above expression (with the basis ) one can arrive at 
\begin{eqnarray}
\hat{\mathcal{H}}_F^\infty\vert\phi_\alpha\rangle=E_\alpha\vert\phi_\alpha\rangle,
\end{eqnarray}
with an infinite time average matrix operator $\tilde{\mathcal{H}}_F^\infty$ defined as \begin{align}\label{Eq8}
	\hat{\mathcal{H}}_F^\infty=\begin{pmatrix}\ddots & \ddots & \ddots & \ddots&\ddots &\ddots & \ddots\\\ddots &  \hat{\mathcal{H}}_{0}^{-2} & \hat{V} & 0 & 0 & 0 &\ddots\\\ddots & \hat{V}^\dagger & \hat{\mathcal{H}}_{0}^{-1} & \hat{V} & 0 & 0 & \ddots\\\ddots &  0 & \hat{V}^\dagger & \hat{\mathcal{H}}_{0}^{0} & \hat{V} & 0 &\ddots\\ \ddots & 0 & 0 & \hat{V}^\dagger & \hat{\mathcal{H}}_{0}^{1} & \hat{V} & \ddots\\ \ddots & 0 & 0  &0 & \hat{V}^\dagger  & \hat{\mathcal{H}}_{0}^{2} & \ddots\\ \ddots & \ddots &\ddots  & \ddots & \ddots & \ddots & \ddots \end{pmatrix}.
\end{align} 
Also, we have defined	\begin{align*}
		\hat{\mathcal{H}}^m_0=
		\hbar v_0\begin{pmatrix}\textbf{k}\cdot \boldsymbol{\sigma} & \Delta\,k_+ \boldsymbol{\sigma}_{0}  \\ \Delta^\ast\,k_- \boldsymbol{\sigma}_{0} & \textbf{k}\cdot \boldsymbol{\sigma} \end{pmatrix}
		+\begin{pmatrix}m\,\hbar\omega\,\boldsymbol{\sigma}_{0} & 0 \\ 0& m\,\hbar\omega\,\boldsymbol{\sigma}_{0} \end{pmatrix}		
	\end{align*}with Floquet replicas shifted by $m\hbar\omega$ %here $\omega_r=\omega a_0/v_0$ and
($m=0,\pm1,\pm2...$). %depicts the photon absorption process.
The off-diagonal blocks capture the influence of an external potential due to the radiation which in this case reads as % such as the trap or a superlattice potential  
	\begin{align}
		\hat{V}={e v_0 A_0}\begin{pmatrix} 0 & 1 & 0 & 0\\ 0 & 0 & 0 & 0\\ \Delta & 0 & 0 & 1\\ 0 & \Delta & 0 & 0 \end{pmatrix}.
	\end{align} 
\begin{figure}[t]
	\includegraphics[width=60.5mm,height=60.5mm]{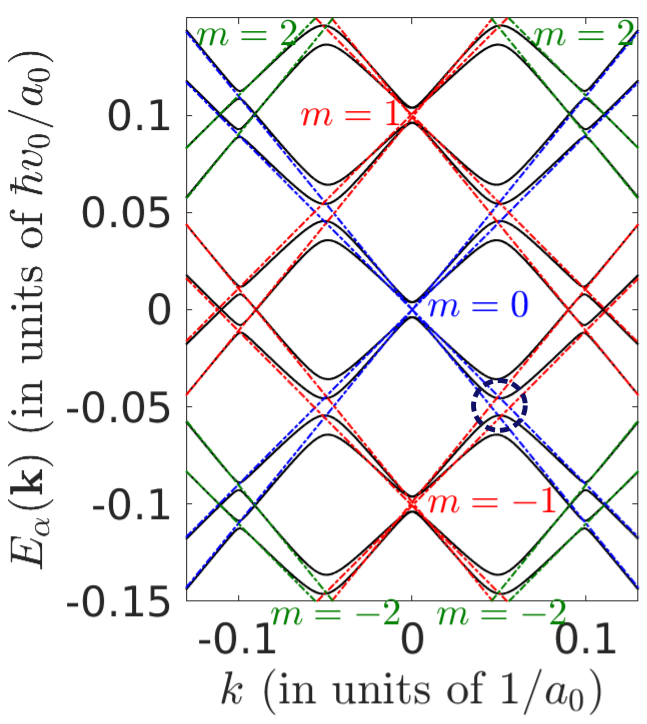}
	\caption{Floquet quasi-energy dispersion with different modes $m$=-2,-1,0,1,2, set forth the quasi-energy gap opening. Here different colors narrate variations in modes. Dotted lines represents the dispersion when the vector potential strength is zero while solid lines exhibit the vector potential influenced dispersion.   }
	\label{Fig3}
\end{figure}
 it is important to mention that only the meaningful choice of quasienergy mode $m$ limits the dimension by  $(8m+4)$ instead of the infinite dimensionality.	In our analysis, we have considered $m=0,\pm1,\pm2$ and the resulting dispersion has been shown in Fig. [\ref{Fig3}]. A considerable simplification can be achieved by adopting Floquet modes with $m=0,-1$ to portray the degeneracy among different modes. So, the truncated Hamiltonian projected onto $m=0,-1$ space looks like 
\begin{align}
\hat{\mathcal{H}}_{\mathcal{F}}=\begin{pmatrix}\hat{\mathcal{H}}_{0}^{-1} & \hat{V} \\ \hat{V}^\dagger & \hat{\mathcal{H}}_{0}^{0} 
		\end{pmatrix}
	\end{align}
For $A_0=0$, among eight eigenenergies, three quasienergies -$\omega/2$, $\mp\omega(1\pm\Delta)/2$ are degenerate at $\textbf{k}=$ $\omega/2(1\pm\Delta)$ and $\omega/2$ as shown the black circled area in Fig.[\ref{Fig3}]. When the finite value of potential strength is turned on, a dynamical gap opening happens at quasi-energy value $m\omega_r/2$ and more significantly this perturbation with $ea_0A_0/\hbar=0.02$ lifts the degeneracy at all the crossing point along with $\Gamma$-point. These gaps lead to the generation of massive fermions with detailed discussions in the next section. %In Fig[\ref{Fig3}(b)], we have also enumerated the Density Of State as a function of energy E, which can be expressed as \begin{eqnarray}
	%	\rho(E)=\frac{1}{4\pi^2}\int{\delta[E(k)-E(k_0))] \,d^2k}
%	\end{eqnarray}  
%	this avails the occupancy and the availability of all these optically driven states %within a particular frequency range. The density of the state near the Fermi level is %directly related
%	to the energy levels of the states near the corners of the Kek-Y superlattice
	%Brillouin zone.

\textbf{{Existense of Valley selective Circular Dichroism and non-trivial topological Invariant number  }}: One of our motives is to check whether a non-equilibrium perturbation can induce a topological phase in Kek-Y distorted graphene. Using degenerate perturbative approach\cite{effective_H1,effective_H2}, the construction of an effective Hamiltonian $
	\hat{\mathcal{H}}_{\rm eff}=\sum_{l=0}^\infty \hat{\mathcal{H}}_{\rm eff}^l $ provides the best way to implement the purpose. It enables to give an accurate description towards the longer time scale dynamics than the driving period $T=2\pi/\omega$ so that the system can be treated in a stroboscopic fassion\cite{kitagawa}. Now lets discuss a little about the range of frequency to disturb a system for a long driven period. When the driving frequency $\omega$ is smaller compared to the  bandwidth $\Sigma$ (which ascertains the range of the quasi-energy eigenstates over which a periodically driven quantum system is distributed), single-particle can only be excited through high-order multiphoton $(m\hbar\omega)$ absorption processes. The optimal order of this excitation process $m$ is governed by the ratio $\Sigma/m\hbar\omega$. Besides another qualitative regime with the approximation $\Sigma<<\hbar\omega$ suggests that one particle is not capable enough to absorb a single photon and getting excited to the higher energy state. In fact this process involves a large order rearrangement of particles so that the rate to perturb the high-energy degrees of freedom can be exponentially suppressed. Thus high-frequency limit\cite{Highfreq} endeavors nothing but the process involving only a single photon\cite{singleph} instead of multiphoton phenomena and dictates the form of truncated effective Hamiltonian as
	
\begin{eqnarray}
\hat{\mathcal{H}}_{\rm eff}=\hat{\mathcal{H}}_0+\frac{[\hat{\mathcal{H}}_{-1},\hat{\mathcal{H}}_{+1}]}{\hbar\omega}
	\end{eqnarray}where $\hat{\mathcal{H}}_{\pm}$ by Fourier transformation as follows
	\begin{eqnarray*}
		\hat{\mathcal{H}}_{\pm}=\frac{1}{T}\int_{0}^{T} \hat{\mathcal{H}}_0(k,t) e^{\mp i\omega t}\,dt.
	\end{eqnarray*} 
Due to the irradiation, this elegant technique modifies the tailored Hamiltonian  by the commutation
	$[\hat{\mathcal{H}}_{-1},\hat{\mathcal{H}}_{+1}]/\hbar\omega=\hat{S}_z^\zeta$  which is given by
	\begin{eqnarray*}
		 \hat{S}_z^\zeta=\zeta\begin{pmatrix}
			\eta_0-\eta_\tau & 0& 0& 0\\0 & -\eta_0-\eta_\tau &0 &0 \\ 0 & 0 & \eta_0+\eta_\tau & 0\\ 0 & 0 & 0 & -\eta_0+\eta_\tau
		\end{pmatrix}
	\end{eqnarray*}
where $\eta_0=e^2 A_0^2 v_0^2/\hbar\omega$ and $\eta_\tau=\Delta^2 \eta_0$ which are quadratic in Fermi velocity ($v_0$). This photon-dressed mass term $\hat{S}_z^\zeta$ changes its sign while polarization switching of CPL takes place (from $\zeta=+$ to $\zeta=-$ and vice-versa). %{\color{blue}The moment it alters the sign, broken Time Reversal Symmerty is realized in Kek-Y Superlattice.} 
The corresponding eigenvalues of $4\cross4$ effective Hamiltonian are as follows \begin{eqnarray}
\scalemath{1}{E_{\lambda\gamma}^{\rm eff}({\bf k})=\gamma(\sqrt{\eta_0^2+\hbar^2 v_0^2 k^2}+\lambda \sqrt{\eta_\tau^2+\hbar^2 v_\tau^2 k^2})}
\end{eqnarray}
Not only that the radiation induces a topological gap, it is also evident from the above equation that the valley degeneracy at the $\Gamma$-point is lifted. The energy gap  can be easily worked out to be %, it is 	By looking at this expression and the dispersion shown in Fig[\ref{Fig4}a], It appears to us that at k=0 point there is no more degeneracy. Now, the energy of quasiparticles depends on the momentum as well as the valley pseudospin. This leads to open a finite gap 
$E_g^\lambda=2(\eta_0+\lambda\eta_\tau)$ showing that it is tunable by the external influence of the vector potential strength $A_0$ associated with CPL. % between two successive cones at the $\Gamma$ point and more interestigly, it now seems possible to tune the gap by  
%With the contrary of the concentric Dirac cones at $\Gamma$ point in non-radiated Kek-Y structure, the resonant coupling of two valley points in any value of momentum space within Brillouin zone can not be sustained by the high-frequency drive. 
As a direct consequence, the degeneracy between the two Dirac cones is lifted (in either of the conduction and the valence band) by $2\eta_\tau$ at the $\Gamma$- point (as shown in Fig. [\ref{Fig4} a)]). These two distinguishable cones suggest we can optically tune the valley degree of freedom in Kek-Y superlattice appropriately. There is a way to reconcile a degree of freedom with the valley pseudospin by calculating the Berry curvature. So, the geometry of Berry phase angle is invoked desirously to describe
the global phase evolution of a complex vector traversed around
a path in its vector space. The Berry connection associated with the eigenenergy $E_{\lambda+}^{\rm eff}(\textbf{k})$ can be defined as\cite{alphat3}\begin{eqnarray}
		\Pi_{\lambda+}(\textbf{k})=i \langle\Psi_{\lambda+}(\textbf{k})|\grad_\textbf{k}|\Psi_{\lambda+}(\textbf{k})\rangle
	\end{eqnarray}
	where $\Psi_{\lambda+}$=$(-\Psi_{BK^\prime}\,\Psi_{AK^\prime}\,\Psi_{AK}\,\Psi_{BK})^T$ and the four components spinor implies about the amplitudes on sublattices A and B for different valleys $K$ and $K^\prime$. The corresponding Berry curvature for  photo-influenced eigenstates is given by \begin{eqnarray}
		\mathcal{B}_{\lambda+}(\textbf{k})=[\grad_\textbf{k}\cross\boldsymbol{\Pi}_{\lambda+}({\bf k})]_z
	\end{eqnarray}
	
 This quantity exhibits  divergence at the $\Gamma$ point and rapidly vanishes away from the center where the singularity arises as depicted in Fig. [\ref{Fig4} b)]. The breaking of TRS leads to the emergence of valley-resolved Chern numbers in photo-illumined Kek-Y layer which we have ellaborated in the following. It is convenient for our analysis to assign the chirality to $K\;(K^\prime)$ valley traverse in a counterclockwise (clockwise) manner with pseudospin up (down).  A non-vanishing Chern number with a certain chirality actually illustrates that the contribution comes only from one valley while the other valley remains inactive under the influence of left/right CPL. This valley dependent assymetry in Berry curvature elucidates the opposite anomalous velocity of charge carriers depending on which valley they are sited. In Fig. [\ref{Fig4} b)], we have shown the response of the Berry curvature of the $K$ and $K^\prime$ states when the polarization state of the radiation is changed from right-handed CPL (RCPL) and in Fig. [\ref{Fig4} c)] left-handed CPL (LCPL). It can be seen that changing in polarisation direction actually leads to flipping of the pseudospin and switches the states between $K$ and $K^\prime$. Thus the different valley-resolved responses towards RCPL and LCPL signal the possibility to achieve a valley-selective Chern number which can direct us one step closer towards the efficient applicability in valleytronics. To further manifest the above observation, we calculate the winding numbers by integrating over the Berry curvature, known as Chern number or topological invariant number\cite{alphat3}\begin{eqnarray}
		\mathcal{C}_{\lambda}=\frac{1}{2\pi}\int_{FBZ}\hat{\textbf{z}}\cdot[\grad_\textbf{k}\cross\boldsymbol{\Pi}_{\lambda+}]\,\textbf{dk}
	\end{eqnarray}
\begin{figure}[t]
 	\includegraphics[width=35.5mm,height=42.5mm]{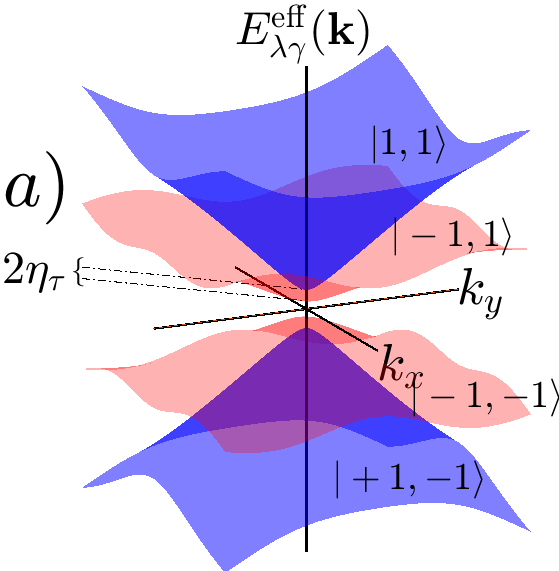}
 	\includegraphics[width=76.5mm,height=75.5mm]{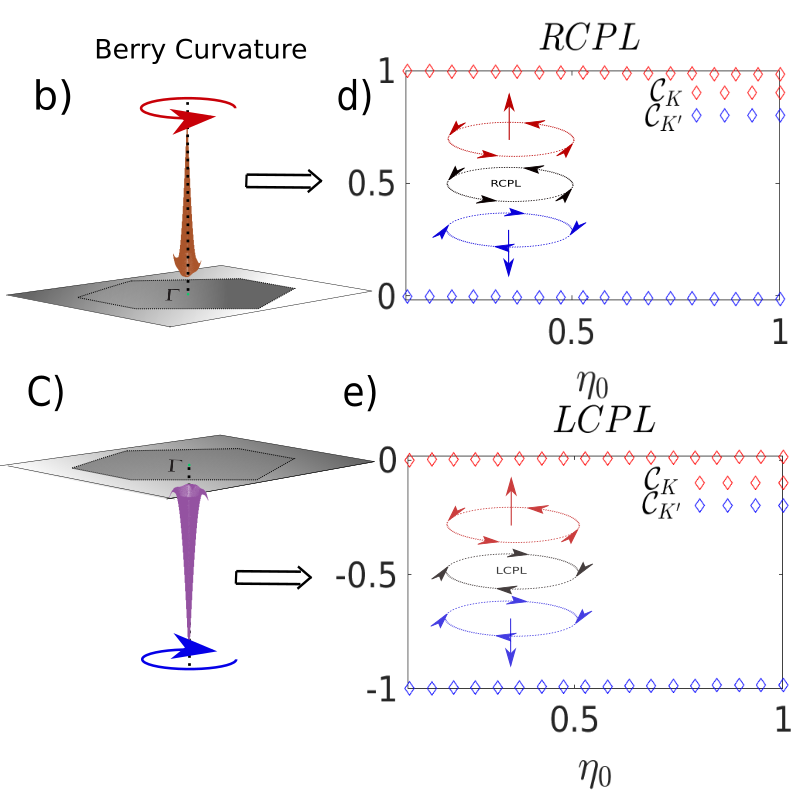} 	
 	\caption{(a) Low energy Dispersion of Photon-dressed Kek-Y  graphene with the gap opening at $\Gamma$-point and breaking the valley degeneracy (b,c) Pictorial  Depiction of the Berry Curvature reveals the valley-selective Circular Dichroism corresponding to RCPL \& LCPL-influenced Floquet state $|\lambda,\gamma\rangle$ respectively (d,e) Valley resolved topological invariant No.($\mathcal{C}_K,\mathcal{C}_{K^\prime}$) with the variation in vector potential strength$(\eta_0)$ associated with RCPL \& LCPL respectively. Here red and blue arrow decode the valley-pseudospin up($\uparrow$) and down($\downarrow$) respectively.}
 	\label{Fig4}
\end{figure}	

In Fig. [\ref{Fig4} d)], $\mathcal{C}_K=1$ unfolds that RCPL drives Kek-Y superlattice only for K-valley and simultaneously $K^\prime$ valley makes no contribution as $\mathcal{C}_{K^\prime}=0$. While on the other hand LCPL alters the number as $\mathcal{C}_K=0$ and $\mathcal{C}_{K^\prime}=-1$ which shows that LCPL excites electrons only in the $K^\prime$ valley keeping silent the charge carrier in the $K$ valley as shown in Fig. [\ref{Fig4} e)]. Thus we have found out that the right (left)-circularly polarized light creates open channels for $K\; (K^\prime)$ valley, leading to optically controlled valley polarisation. A reliable explanation to valley-selective CD is that when the pseudospin winding of valley and the polarization of driving field ( RCPL or LCPL) have the same chirality, the valley resolved Chern number becomes non-trivial $(\mathcal{C}_{K\text{ or }K^\prime}=\pm1)$. When the chiralities are opposite, the winding vanishes, thus the Chern number. Thereby, these outcomes confirm the existence of the Valley Quantum Anomalous Hall (VQAH) states\cite{floqvalley} in illuminated Kek-Y structured superlattice. So, our findings pave a way to achieve optically switchable valley-filter in Kek-Y graphene superlattice. 
 
\section{Degree of Valley-resolved optical polarisation}\label{sec4}
We will now provide a consolidation to understand the valley-resolved optical polarisation due to CPL which is desired for valleytronics application. The inter-band transitions from  valence band to the conduction band is the key to extract the degree of valley polarisation\cite{CPLMoS,CDsilicene,natureMoS}. In order to have the analytical expression for optical transition, we have made an assumption $\eta_\tau<<\eta_0$ as $\eta_\tau\propto\Delta^2$ and obtained the energy eigenvalues  as \begin{eqnarray}
E(\lambda,\gamma)=\gamma(\lambda k v_\tau \hbar+\sqrt{\eta_0^2+k^2v_0^2\hbar^2})
	\end{eqnarray} The ascending order of these energies is defined as $E_{1, -1}<E_{-1, -1}<E_{-1, 1}<E_{1, 1}$. The coupling strength with the optical fields associated with CPL is given by
	$P^\zeta(\textbf{k})=P_x^\zeta+i\zeta P_y^\zeta$ where $P_{x,y}^\zeta=\langle{\Psi_{\lambda^\prime,+}}|\frac{1}{\hbar}\frac{\partial H_eff}{\partial k_{x,y}}|{\Psi_{\lambda,-}}\rangle$\cite{CPLMoS,CDsilicene,natureMoS}. The transition matrix element near $\Gamma$ point in the limit $\eta_\tau\rightarrow 0$ goes as
	\begin{eqnarray}\label{eq16}
	|P^\zeta_{\lambda\rightarrow\lambda^{\prime}}(\textbf{k})|^2=
	\begin{cases}
   0 & \text{if}\ \lambda=\lambda^\prime\\
   v_0^2\left(1+\lambda\zeta\frac{\eta_0}{\sqrt{\eta_0^2+k^2v_0^2\hbar^2}}\right)^2 & \text{if}\ \lambda\neq\lambda^\prime		
    \end{cases}
	\end{eqnarray} where $\lambda, \lambda^\prime$ encodes the initial, final cone index. It is important to mention that all the intra-cone transitions are forbidden in the presence of light of any polarization according to the obtained optical-selection rule in Eqn. (\ref{eq16}). The inter-cone optical transitions i.e. E(1,-1)$\rightarrow $E(-1,1) and E(-1,-1)$\rightarrow$E(1,1) do exist and are coupled to only LCPL and RCPL respectively. Previously discussed valley-dependent CD with non-zero Chern numbers confirms that LCPL-coupled transition occures due to the excitation of $K$-valley fermions and the contribution in case of RCPL-coupled transition comes from the electron in $K^\prime$-valley. So, now the valley-resolved optical polarisation can be defined as\cite{CPLMoS,CDsilicene,natureMoS}
	\begin{eqnarray}
\mathcal{P}(\textbf{k})=\frac{|P^-_{\vert\lambda,+\rangle\rightarrow\vert\lambda^{\prime},-\rangle}(\textbf{k})|^2-|P^+_{\vert\lambda,+\rangle\rightarrow\vert\lambda^{\prime},-\rangle}(\textbf{k})|^2}{|P^-_{\vert\lambda,+\rangle\rightarrow\vert\lambda^{\prime},-\rangle}(\textbf{k})|^2+|P^+_{\vert\lambda,+\rangle\rightarrow\vert\lambda^{\prime},-\rangle}(\textbf{k})|^2}
	\end{eqnarray}
	This quantity narrates the difference between the absorption of LCPL and RCPL($\zeta=\pm$) by $K^\prime$, $K$ valley respectively, normalized by the
	total absorption at each k-point in the momentum space. Our rigorous calculation suggests that the analytical expression for $\mathcal{P}(\textbf{k})$ is as follows
	\begin{eqnarray}
	\mathcal{P}(\textbf{k})=\pm\frac{2 \eta_0 \sqrt{\eta_0^2+k^2 v_0^2 \hbar^2}}{2 \eta_0^2+k^2 v_0^2 \hbar^2},
	\end{eqnarray}
	which precisely enumerates and predicts the optical valley pollarisation at $\Gamma$-point as $\mathcal{P}(\textbf{k}\rightarrow0)=\pm1$. Here '+' and '-' corresponds to $\vert1,1\rangle\rightarrow\vert-1,-1\rangle$ and $\vert-1,1\rangle\rightarrow\vert1,-1\rangle$ transitions respectively. This degree of polarisation starts to approach zero value as k-value drifts away from high-symmetric $\Gamma$ point. As a direct consequence, photoemission enables to achieve such kind of complete valley polarisation at $\Gamma$-point. Having non-trivial Chern numbers and valley-polarisation, in the next part of our analysis we can further proceed to infer the appearence of topological transition\cite{floquetopo}. 
\section{Photon enhanced nano-ribbon geometry approach in Kek-Y superlattice   }\label{sec5}
\begin{figure}[t]
	\includegraphics[width=85mm,height=53.5mm]{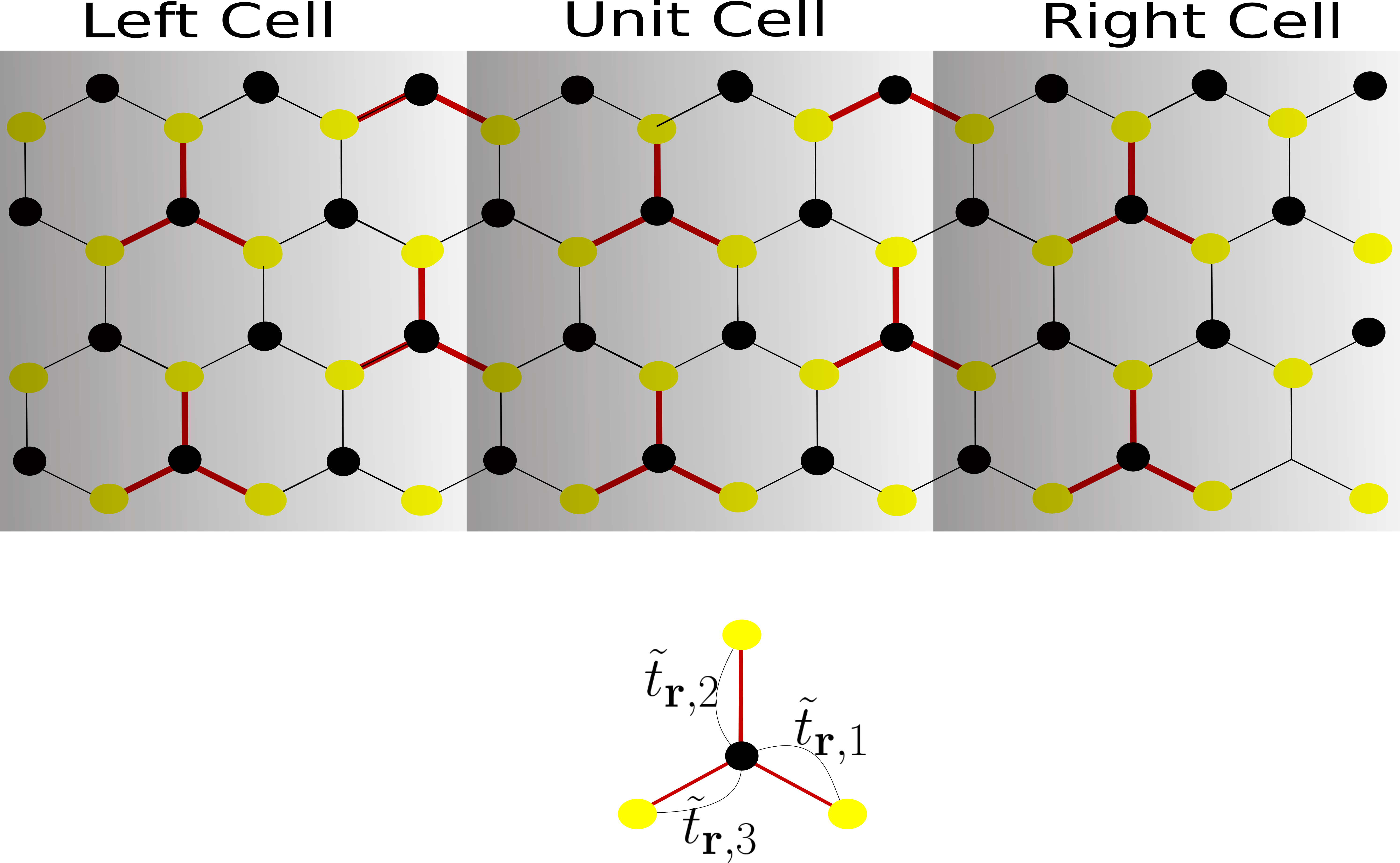}
	\caption{Schematic illustration shows the real-space lattice structure of Y-shaped distorted graphene with the space dependent and photo-induced hopping parameter($\tilde{t}_{r,\xi}$).}
	\label{Fig6}
\end{figure}
The landscape of nano-ribbon approach provides a proper framework to study the electronic transport properties in Kek-Y distorted graphene nanoribbon. In our analysis, we have considered a zigzag nanoribbon geometry which holds quite large snatch of carbon atoms sitting on the edge. Theory of bulk boundary correspondence\cite{BBC}in a topological system allows the edges to develop totally unique features which can only exist because of the bulk properties. So, to witness the existence of topological edge states\cite{edge1,edge2,edge3,edge4,edge5} let us start with the Hamiltonian with Tight-Binding approximation \cite{keknano}    
	\begin{equation}\label{eq19}
		\hat{H}_\mathcal{TB}=\hat{\mathcal{E}}+\hat{\Lambda}\, e^{3ik_x }+\hat{\Lambda}^\dagger e^{-3ik_x}.
	\end{equation}
	Here $\hat{\mathcal{E}}$ be the matrix elements denoting the interaction among the sites within the same unit cell, $\hat{\Lambda}(\hat{\Lambda}^\dagger)$ stands for the coupling to  one unit cell with the sites of its right(left) neighboring cell. Explicit expression of $\hat{\mathcal{E}}$ and $\hat{\Lambda}(\hat{\Lambda}^\dagger)$ are shown in the Appendix A.
% where $t_1=1-\Delta_0$ and $1+2\Delta=1+2\Delta_0$
Due to Kekul$\'e$-Y distortion, the chosen unit cell in Kek-Y graphene is three times larger than the pristine graphene unit cell which is essential to unveil the characteristics exhibited by the material as shown in Fig.[\ref{Fig5} a)]. The width of the nanoribbon is decided by $N=20$ iterative cells of the unit cell. 

By solving Eq. (\ref{eq19}), we have shown the dispersion in Fig. [\ref{Fig5} a)].
\begin{figure*}[t]
	\includegraphics[width=181mm, height=60.5mm]{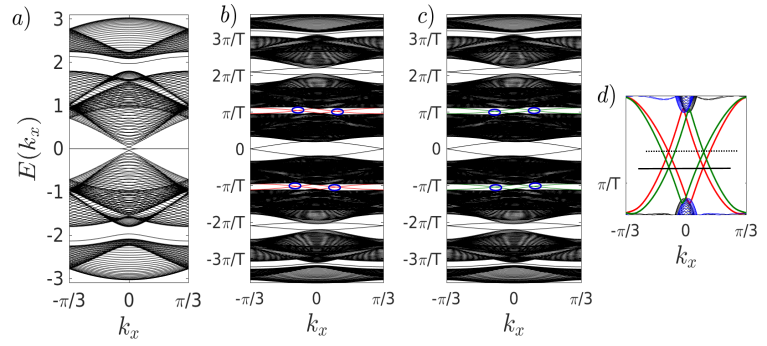}
	\caption{The band diagram of Kek-Y($\nu$=1) zigzag distorted graphene nanoribbon as predicted by tight-binding Hamiltonian\,(a) in the absence of radiation\, (b \& c) when Kek-Y structure is exposed to the RCPL \& LCPL (taking $\hbar\omega=3.2t_0$), which initiates quasi-energy dispersion with periodic chiral edge modes\,(d) zooming in the topological edge states with prominent band crossing. The red/green curves depict counter-propagating edge states corresponding the RCPL \& LCPL respectively.}
	\label{Fig5}
\end{figure*}	
	 Here the overlapping of the highest level of valence band with the lowest energy level of conduction band has been observed throughout the Brillouin zone. Emergence of this flat band at zero-energy level signifies the localized states at the surface consisting Kramer degeneracy\cite{keknano,flatband}. Now, we would like to reveal how this energy dispersion evolves with the driven field showcasing  the reformulated effective difference Hamiltonian as
	\begin{equation}
		\hat{H}_\mathcal{TB}(t)=\hat{H}_s+\hat{\tilde{\Lambda}}_s\, e^{3ik_x a}+\hat{\tilde{\Lambda}}_s^\dagger e^{-3ik_x a}
	\end{equation}   
	here $\hat{H_s}, \hat{\tilde{\Lambda}}_s$ are properly defined in the Appendix B. Also, the position dependent hopping parameter  ($t_{\textbf{r},\xi}$) innovates itself as $\tilde{t}_{\textbf{r},\xi}=t_{\textbf{r},\xi} \,\exp\left[i \frac{2\pi}{\phi_0}\int_{\textbf{r}}^{\textbf{r}+\textbf{q}_\xi}\textbf{A}.\textbf{dr}\right]$
	and  $\phi_0$ is magnetic flux quanta. Jacobi-Anger expansion $e^{iz \sin{\theta}}=\sum_{s=\infty}^{-\infty}\mathcal{J}_s(z) e^{is\theta}$ with the s-th order cylindrical Bessel function $\mathcal{J}_s(z)$ architects the photon dressed tight binding Hamiltonian in such an intuitive way providing
	\begin{eqnarray*}
	\tilde{t}_{\textbf{r},1}=t_{\textbf{r},1}\sum_{s=-\infty}^{\infty}\mathcal{J}_s(-z) e^{is(\omega t-\pi/3)}\\
			\tilde{t}_{\textbf{r},2}=t_{\textbf{r},2}\sum_{s=-\infty}^{\infty}\mathcal{J}_s(z) e^{-i s (\omega t+\pi/3)}
		\\
		\tilde{t}_{\textbf{r},3}=t_{\textbf{r},3}\sum_{s=-\infty}^{\infty}\mathcal{J}_s(z) e^{is\omega t}
		\end{eqnarray*}
where the dimensionless quantity $z=2\pi A_0 a/\phi_0$ brings forth the strength of circularly polarised light. Following the similar approach as per the non-radiated case, we arrive at infinite time average on-site Floquet Hamiltonian with the form of
	\begin{align}
		\hat{H}_\mathcal{F}=\begin{pmatrix} \ddots & \ddots & \ddots & \ddots & \ddots & \ddots & \ddots \\\ddots & \hat{H_1}^\dagger & \hat{H}_0-\hbar\omega& \hat{H}_1 & \hat{H}_2 & \hat{H}_3 & \ddots \\\ddots & \hat{{H_2}}^\dagger & \hat{{H_1}}^\dagger & \hat{H}_0 & \hat{H}_1 & \hat{H}_2 & \ddots \\\ddots & \hat{{H_3}}^\dagger & \hat{{H_2}}^\dagger & \hat{{H_1}}^\dagger & \hat{H_0}+\hbar\omega & \hat{H}_1 & \ddots \\\ddots & \ddots & \ddots & \ddots & \ddots & \ddots & \ddots    \end{pmatrix}
	\end{align}

%In a similar way, we can also obtain the coupling Hamiltonian $\tilde{\Lambda}$.	
In order to yield energy dispersion, we gather all the possible eigenvalues of Floquet infinite dimensional Hamiltonian $\hat{H}_\mathcal{F}$ carring all the modes $m$. Analysing the outcomes we can make definite conclusion about the evolution of edge states as a consequence of driven field.

\textbf{Uprising Valley resolved Chiral Edge States by Circularly Polarized Light} :
In this section, our analysis aligns towards the topological aspects of Floquet Hamiltonian which provides hints to predict the fate
of edge states. In Fig.[\ref{Fig5} b,c)], we therefore consider the nano-ribbon geometry and plot the energy dispersion accounting for the influence of both the polarisation(RCPL \& LCPL). This figure elucidates that the periodicity in CPL stimulates the energy dispersion to replicate in high frequency regime. But it is clearly noticable that the spectral width starts to decrease as we move apart from $m=0$ mode. We have used the radiation having energy $\hbar\omega=4 t_0$ in the unit of eV with photon modes $m=0,\pm1,\pm2$ and $z=0.8$. Let's focus on one replica where the vivid band crossing of two distinct states taking place. In this figure, two pair of edge modes(blue-circled) counterpropagate through first Floquet-Brillouin zone. Thus the non-trivial winding arises in one driving cycle[$-\frac{\pi}{T}\,\frac{\pi}{T}$] and complete the quasi-energy cycle connecting the top of the upper Floquet band to the bottom of the lower Floquet band. As per our calculation in section \ref{sec3}, Chern number comes out to be non-trivial which shows the influencial photo-driving invokes topological effect into Kek-Y superlattice. Interestingly, this non-zero contribution arises due to the localisation of valley contrast electronic wave-function resides on A/B-sublattice. We have zoomed out the part of the localized edge states in Fig. [\ref{Fig5} c)]. Having had an observation of valley-resolved Chern number $\mathcal{C}_{\lambda\gamma}=\pm1$, it is now clear that under the influence of RCPL, two counterpropagating edge states contribution come from  the charge carriers having pseudospin up in valley $K$  as it always preserves its non-trivial topology with the chirality same as polarisation. Besides, the states in $K^\prime$ valley develop a sense to be canceled out due to its opposite chirality and penetrate into the bulk. Conversely when the superlattice is subjected to LCPL, only state undergoes to the topological phase transition corresponding to $K^\prime$ valley. In Fig. [\ref{Fig5}b,c)] we have exploited the effect of RCPL and LCPL respectively. We have also witnessed that LCPL compels the energy dispersion to alter in such a way that it innovates itself exactly as  a mirror image of Fig. [\ref{Fig5} b)] w.r.t the momentum axis. Now this kind of inversion puts forward the evidence which is strong enough to conclude from our analysis that the state switching is actually taking place with changing of the polarisation.

\section{conclusions and summary}\label{sec6}
We have initiated the study of Floquet dynamics with Kek-Y distorted graphene superlattice. Based on the helicity of non-vanishing Berry curvature}, we can be assured about the pseudospins which are parallel/antiparallel to the momentum associated with $K/K^\prime$ valleys. This valley-momentum locking phenomenon is responsible for breaking the degeneracy between $K$ and $K^\prime$ cones having different Fermi velocities except $\Gamma$ point. But interestingly, we have found out that it is possible to lift the valley degeneracy by $2\eta$ at low momentum region by breaking T symmetry inducing CPL.  The estimated electric field  coming out from our analysis is $60.8\cross 10^9$ V/m and the intensity is I $\approx e^2 A_0^2 \,\omega^2/\hbar \approx 10^{15}$ W/$m^2$. Also, we have obtained valley selective non-trivial Chern number $\mathcal{C}_{K}/\mathcal{C}_{K^\prime}$=+1/-1 which unfolds that the entire valley $K^\prime$ absorbs almost purely left-handed photons whereas $K$ valley responds to right-handed photons. With this we have investigated enough by exploring valley-contrast polarisation to conclude that valley-selective circular dichroism(CD) which obeys optical valley selection rule is achievable in this photo-induced system. In a nutshell, when RCPL(LCPL) is shone over the sample,  a non-equilibrium state is created in both valleys where the charge carrier population accumulate mostly in $K$ ($K^\prime$) valley. As the distinguishable response of both the valleys and valley resolved counterpropagating edge states are deemed appropriate evidence, we have proposed a  beneficial way out in this work to achieve VQAH state in photodriven Kek-Y distorted graphene superlattice. So, our investigation holds significant psomise for advancing the capabilities of an optically switchable topological valley filter which  is suitable for valleytronic devices.
	
	\textit{Acknowledgments}: This work is an outcome of
	the Research work carried out under the DST-INSPIRE project DST/INSPIRE/04/2019/000642, Government of India.
	
 \section*{Appendix A}\label{AppB}
 We provide below the explicit form of the on-site potential energy mentioned in section-\ref{sec5} within the same unit cell as
	\begin{eqnarray*}
		\hat{\mathcal{E}}&=& \Bigg[\begin{pmatrix}
			0 &  2\Delta & 0 & 0 & 0 & 0\\ 2\Delta^\ast & 0 & 2\Delta & 0 & 0 & 0 \\ 0 & 2\Delta^\ast & 0 & -\Delta & 0 & 0 \\ 0 & 0 & -\Delta^\ast & 0 & -\Delta &0\\ 0 & 0 & 0 & -\Delta^\ast & 0 & -\Delta\\0 & 0 & 0 & 0 & -\Delta^\ast & 0
		\end{pmatrix}\\\nonumber&+&\begin{pmatrix}
			0 &  1 & 0 & 0 & 0 & 0\\ 1 & 0 & 1 & 0 & 0 & 0 \\ 0 & 1 & 0 & 1 & 0 & 0 \\ 0 & 0 & 1 & 0 & 1&0\\ 0 & 0 & 0 & 1 & 0 & 1\\0 & 0 & 0 & 0 & 1 & 0
		\end{pmatrix}\Bigg]\otimes\mathbb{I}_{2N\times 2N}+\begin{pmatrix}
			0_{6 \times 6} & h\\ h^\dagger & 0_{6\times6} 
		\end{pmatrix}\otimes\mathbb{I}_{N\times N}
	\end{eqnarray*} here h matrix is defined as 
	\begin{eqnarray*}
		\hat{h}=\begin{pmatrix}
			0 & 0 & 0 & 0 & 0 & 0\\ 0 & 0 & 0 & 0 & 1+2\Delta & 0 \\ 0 & 0 & 0 & 0 & 0 & 0 \\ 0 & 0 & 1-\Delta & 0 & 0 &0\\ 0 & 0 & 0 & 0 & 0 & 0 \\1-\Delta & 0 & 0 & 0 & 0 & 0 
		\end{pmatrix}
	\end{eqnarray*}
	and $\mathbb{I}_{N\times N}$ as an identity matrix where N be the number of unit cell. The coupling matrix between the unit cell and it's right cell as shown in Fig.[\ref{Fig6}] can be written as
	\begin{equation*}
	\hat{\Lambda}= \begin{pmatrix}
	0 &  0 & 0 & 0 & 0 & 1-\Delta\\ 0 & 0 & 0 & 0 & 0 & 0 \\ 0 & 0 & 0 & 0 & 0 & 0 \\ 0 & 0 & 0 & 0 & 0 &0\\ 0 & 0 & 0 & 0 & 0 & 0\\ 0 & 0 & 0 & 0 & 0 & 0
	\end{pmatrix}\otimes{I}_{2N\times 2N}
	\end{equation*}

\section*{Appendix B}\label{AppB}
Here we delve into the form of $\hat{H}_s$ and $\hat{\tilde{\Lambda}}_s$ which retaining the complete information of the s-th order cylindrical Bessel function $\mathcal{J}_s(z)$. Our rigorous calculation suggests the definite expressions as following 
\begin{widetext}
\begin{eqnarray*}
	\hat{H}_s&=& \mathcal{J}_s(z)\Bigg(e^{i s\pi/3}\Bigg[\begin{pmatrix}
		0 &  0 & 0 & 0 & 0 & 0\\ 2\Delta^\ast & 0 & 2\Delta & 0 & 0 & 0 \\ 0 & 0 & 0 & 0 & 0 & 0 \\ 0 & 0 & -\Delta^\ast & 0 & -\Delta &0\\ 0 & 0 & 0 & 0 & 0 & 0\\0 & 0 & 0 & 0 & -\Delta^\ast & 0
	\end{pmatrix}+\begin{pmatrix}
		0 &  0 & 0 & 0 & 0 & 0\\ 1 & 0 & 1 & 0 & 0 & 0 \\ 0 & 0 & 0 & 0 & 0 & 0 \\ 0 & 0 & 1 & 0 & 1&0\\ 0 & 0 & 0 & 0 & 0 & 0\\0 & 0 & 0 & 0 & 1 & 0
	\end{pmatrix}\Bigg]+e^{-i s\pi/3}\Bigg[\begin{pmatrix}
		0 &  1 & 0 & 0 & 0 & 0\\ 0 & 0 & 0 & 0 & 0 & 0 \\ 0 & 1 & 0 & 1 & 0 & 0 \\ 0 & 0 & 0 & 0 & 0&0\\ 0 & 0 & 0 & 1 & 0 & 1\\0 & 0 & 0 & 0 & 0 & 0
	\end{pmatrix}\\\nonumber&+&\begin{pmatrix}
		0 &  2\Delta & 0 & 0 & 0 & 0\\ 0 & 0 & 0 & 0 & 0 & 0 \\ 0 & 2\Delta^\ast & 0 & -\Delta & 0 & 0 \\ 0 & 0 & 0 & 0 & 0 &0\\ 0 & 0 & 0 & -\Delta^\ast & 0 & 2\Delta\\0 & 0 & 0 & 0 & 0 & 0
	\end{pmatrix}\Bigg]\Bigg)\otimes\mathbb{I}_{2N\times 2N}+\begin{pmatrix}
		0_{6 \times 6} & \mathcal{J}_s(-z)h\\ \mathcal{J}_s(z)h^\dagger & 0_{6\times6} 
	\end{pmatrix}\otimes\mathbb{I}_{N\times N}.
\end{eqnarray*}
\end{widetext}
and
\hfil
\begin{eqnarray*}
	\hat{\tilde{\Lambda}}_s=\begin{pmatrix}
		\hat{h}_1 & 0_{6\times6}\\0_{6\times6} & \hat{h}_1^\dagger
	\end{pmatrix}\otimes\mathbb{I}_{N\times N}
\end{eqnarray*}with

\begin{eqnarray*}
	\hat{h}_1=\mathcal{J}_s(z)\begin{pmatrix}
		0 &  0 & 0 & 0 & 0 & 0\\0 & 0 & 0 & 0 & 0 & 0\\0 & 0 & 0 & 0 & 0 & 0\\0 & 0 & 0 & 0 & 0 & 0\\0 & 0 & 0 & 0 & 0 & 0\\(1-\Delta) e^{is\pi/3} & 0 & 0 & 0 & 0 & 0
	\end{pmatrix}
\end{eqnarray*}
 
All these parameters are calculated considering the impact of photo-illumination.

\end{document}